\documentclass[letterpaper, twocolumn, superscriptaddress, showpacs, preprintnumbers, amsmath, amssymb,longbibliography, nofootinbib]{revtex4-1}
\usepackage[pdftex]{graphicx}
\usepackage{caption}
\usepackage{subfigure}
\usepackage{dcolumn}
\usepackage{bm}
\usepackage{color}
\usepackage{hyperref}
\usepackage{adjustbox}
\usepackage{color}
\usepackage{float}
\usepackage{framed}
\usepackage{esvect}
\usepackage{amsfonts,amssymb,amsmath,hyperref,hypcap,enumerate}
\usepackage{wasysym}
\usepackage{slashed}
\usepackage{tikz}
\usepackage{amsmath,amssymb,bm,graphicx,dsfont}
\usepackage[latin9]{inputenc}
\usepackage{amsmath}
\usepackage{slashed}
\usepackage{dsfont}
\usepackage{amstext}
\usepackage{amssymb}
\usepackage{amsbsy}
\usepackage{amsthm}
\usepackage{epsfig}
\usepackage{tikz}
\usepackage{bbm}
\usepackage{hyperref}
\usepackage{color}
\usepackage{multirow}
\usepackage{ulem}
\usepackage{ulem}
\usepackage{soul}
\usepackage[T1]{fontenc}
\newcommand{\be}{\begin{equation}}
\newcommand{\ba}{\begin{align}}
\newcommand{\ee}{\end{equation}}
\newcommand{\bea}{\begin{eqnarray}}
\newcommand{\eea}{\end{eqnarray}}
\newcommand{\beq}{\begin{equation}}
\newcommand{\eeq}{\end{equation}}
\newcommand{\beqn}{\begin{eqnarray}}
\newcommand{\eeqn}{\end{eqnarray}}

\renewcommand{\vec}[1]{{\bf #1}}

\begin{document}
\title{Translation-enriched $Z_2$ spin liquids and topological vison bands: Possible application to $\alpha$-RuCl$_3$}
\author{Xue-Yang Song}
\affiliation{Department of Physics,  Massachusetts Institute of Technology, MA 02139, USA}
\author{T. Senthil}
\affiliation{Department of Physics,  Massachusetts Institute of Technology, MA 02139, USA}
\date{\today}

\begin{abstract}

Inspired by experiments on the magnetic field induced phases of the spin-orbit coupled $2d$ Mott insulator $\alpha$-RuCl$_3$, we study some general aspects of gapped $Z_2$ Quantum Spin Liquids (QSL) enriched by lattice translation symmetry. We show that there are $12$ distinct such phases with different implementations of translation symmetry. In some of these phases the vison excitations of this QSL may form topological Chern bands.  We explore a phenomenological description of a putative  $Z_2$ QSL 
as a candidate ground state at intermediate magnetic fields in $\alpha$-RuCl$_3$. This state has broad continuum spectra in neutron scattering, a ``bosonic'' thermal Hall signal that goes to zero at zero temperature, and is naturally proximate to a zigzag magnetic ordered state, all of which are also seen in $\alpha$-RuCl$_3$. On general grounds continuum scattering will also be seen at multiple points in the Brillouin zone in this state.
%We discuss the possibility of a vison contribution to thermal hall effects in a $Z_2$ spin liquid that breaks time reversal($T$). We classify the symmetry-enriched $Z_2$ topological(SET) order with translations.  Among the $9$ SET classes, we identify and prove classes where a chern band for visons is forbidden. We list the principles and construct fully frustrated vison models, on square, honeycomb and Kagome lattices, that yield a Chern band and hence nontrivial thermal hall effects.  
%Motivated by a recent measurement of thermal hall conductivity $\kappa_{xy}$ in $\alpha-$RuCl$_3$ fitted by a bosonic Chern band \cite{Ong2021}, we propose a vison model where one translation permutes $e,m$ particles.  The confinement of such $Z_2$ order breaks translation, which may lead to the zigzag order of $\alpha-$RuCl$_3$ at small magnetic fields. The vison model holds Chern bands and nontrivial thermal hall conductivity, and connects naturally to spin-wave descriptions in the high-field regime. The distinct symmetry fractionalization predicts enhanced periodicity in neutron scattering.
\end{abstract}
\maketitle

The search for a Quantum Spin Liquid (QSL)  state in electronic materials has led to intensive study of $\alpha$-RuCl$_3$ in the last few years\cite{broholm2020quantum,takagi2019concept,trebst2022kitaev,hermanns2017physics,winter2017models}. This is a layered material with spin-orbit coupled $J = 1/2$ moments arranged on a honeycomb lattice in each layer.  $\alpha$-RuCl$_3$ is deemed a `Kitaev material'\cite{jackeli2009mott} where the dominant interactions yield a Hamiltonian whose exact solution\cite{kitaev_2006} (by Kitaev) reveals a quantum spin liquid phase. Like many other Kitaev materials, the moments  of $\alpha$-RuCl$_3$ order magnetically (into a zigzag pattern) at low temperature due to the effect of other interactions beyond the exactly solvable model. The magnetic ordering is suppressed by application of a modest in-plane field of about $B_c \approx 7 T$. Attention has thus shifted to the possibility of a QSL phase at intermediate fields just beyond $B_c$ before the onset of the expected phase smoothly connected to the fully spin-polarized paramagnet. 

In the region between $B_c \approx 7 T$ and about $9 T$, a number of phenomena have been found which hint at a possible QSL state. Most intriguing is the observation of a large thermal Hall signal even when the magnetic field is strictly in-plane. Early measurements suggested a quantized thermal Hall conductance\cite{matsuda1,takagi} consistent with that expected for a specific field-induced non-abelian QSL state (known as the Ising anyon state) in the ideal Kitaev model. Subsequent experiments\cite{Ong2021} however report an unquantized, strongly temperature dependent, thermal Hall effect that at higher fields could be understood as coming from a topological magnon found in spin wave calculations of the fully polarized state\cite{magnon_ybkim,magnon_moessner,magnon_joshi,moore18}.The Hall signal is seen when the field is applied along the [$11\bar 2$] direction in the coordinate system of the underlying octahedra\cite{kee_nc} (i.e.the a axis in fig\ref{fig:alphamodel}(a), perpendicular to a honeycomb bond) , and is absent when the field is along [$1\bar 10$], parallel to the bonds (the b axis). This is consistent with the presence of a $\pi$ rotation symmetry $C_{2b}$ in the latter case. For a [$11\bar 2$] field, the only symmetry that is preserved is lattice translation (and $C_{2b}\mathcal T$ where $\mathcal T$ is time reversal) and a non-zero thermal Hall effect is not forbidden.

The observed\cite{Ong2021} suppression of the thermal Hall effect with decreasing temperature calls into question the scenario of the field induced non-abelian Ising anyon state of the ideal Kitaev model. Nevertheless spectroscopy measurements do not observe any sharp peaks but just a broad continuum scattering at fields just beyond $B_c$ \cite{neutron_17,neutron2018,neutron_balz,loidl_thz} keeping hopes alive for some kind of field-induced QSL, even if not the Ising anyon state. Both neutron\cite{neutron2018,neutron_balz} and thermodynamic measurements\cite{heat2017,heat2022}  are consistent with an excitation gap, and thus the possibility of a gapped spin liquid at intermediate $B$.  Microscopically a realistic model\cite{jackeli2009mott,rau2014generic,plumb_gamma} of $\alpha$-RuCl$_3$ must  supplement the Kitaev term with other terms, including Heisenberg and anisotropic interactions which are not too small. The phase diagram of such a model in an in-plane field  is not known reliably, despite important numerical studies on small system sizes\cite{kee_nc}. Available results show that the exactly soluble spin liquid state is destabilized by these other terms; instead  magnetically ordered states and indications of other QSL states are found. Thus the existence of a QSL ground state in $\alpha$-RuCl$_3$ remains a tantalizing, but hardly settled, possibility. 

Motivated by this state of affairs, we consider a simple gapped $Z_2$ QSL, enriched by lattice translation but no other global symmetries, as a candidate ground state for $\alpha$-RuCl$_3$ in a generic field direction (including [$11\bar 2$]) \footnote{$C_{2b}\mathcal T$ is broken by $B$ along generic directions but preserved by $B\parallel$[$11\bar 2$]. We ignore $C_{2b}T$ in most of our discussion and account for it in a phenomenological model presented later.}.   We explore the idea that the thermal hall signal may originate from topological bands of fractional excitations called visons of such a QSL. Ref \cite{rosch_vison} considered thermal Hall effects from  visons (really non-abelions) of the Ising anyon state from perturbing the Kitaev model, where visons form Chern bands leading to a large thermal Hall signal at intermediate temperatures.  Eventually though, at the lowest  $T$, the quantized Majorana fermion thermal Hall conductivity will result. Here we instead study ordinary $Z_2$ QSLs with no quantized thermal Hall conductivity in the low-$T$ limit.  

We first classify all such translation enriched gapped $Z_2$ QSLs in $2d$. We find $12$ distinct translation symmetric $Z_2$ QSLs. In some of these  translation acts by permuting topologically distinct quasiparticles.  We identify general principles that allow visons to form a band with nonzero Chern number. These formal considerations will be helpful in interpreting experiments and numerics on Kitaev materials and their models. As an application,  we explore a phenomenological description of $\alpha$-RuCl$_3$ in an intermediate strength  [$11\bar 2$] field as a particular $Z_2$ QSL where translation along one direction permutes topological quasiparticles. This state is naturally proximate to the zigzag orders at low fields. It accommodates vison Chern bands resulting in a thermal hall signal consistent with observations. We comment on the relationship with the thermal hall effect at larger $B$ which is likely  due to topological magnons. This QSL will have continuum scattering in neutron spectroscopy with a characteristic periodicity within the Brillouin zone as an experimental signature. The possibility of a non-Kitaev $Z_2$ QSL has been explored before\cite{chen_19} with the full space group symmetries of honeycomb lattices and time reversal as a candidate for a `proximate' \emph{zero-field} spin liquid in RuCl$_3$.

\emph{Translation-enriched $Z_2$ QSL classification}  $Z_2$ QSLs are topologically ordered and have two distinct bosonic gapped anyon excitations  $e,m$   with $\pi$ mutual statistics in a $2D$ system. The fusion rules are 
\begin{eqnarray}
\label{emfusion}
e\times e=1,m\times m=1,
e\times m=\epsilon
\end{eqnarray}
with the $\epsilon$ a fermion (with $\pi$ mutual statistics with either $e$ or $m$).   
Consider the realizations of such QSLs on a lattice with symmetry under elementary translations along $2$ primitive lattice vectors $T_{1,2}$, the angle between which are taken to be smaller than or equal to $90^o$. %We note that the requirement that two $e(m)$ fusion into a physical excitation and have trivial translations is automatically satisfied regardless $T_1T_2T_1^{-1}T_2^{-1}=\pm 1$.
Despite the  tremendous progress in classifying symmetry enriched topological ordered(SET) states\cite{hermele, levin,barkeshli,qi_z2set,rao2021theory} from diverse approaches, for translation symmetries, only partial results are available in the published literature.  Here we use a physical approach to find all distinct translation enriched $Z_2$ QSLs in a $2D$ system.

In classifying SET states, it is convenient to ask about the action of the symmetry on the topological quasiparticle excitations. There are two distinct situations which require separate treatments. First, the symmetry action does not permute any anyons. Then we can discuss the action of the symmetry on each anyon species separately. This action may be projective, {\em i.e}, the symmetry may be fractionalized by the anyon. Distinct SET states are obtained by assigning distinct projective representations to the various anyons together with some consistency requirements. The second case - the symmetry action includes a permutation of anyons. The anyons being permuted must have identical topological properties for this to be allowed. For the $Z_2$ QSL, this means that some translation could act by interchanging the $e$ and $m$ particles. A number of examples have been constructed of such anyon exchange by translation. Examples include the strong bond limit of the Kitaev model\cite{kitaev_2006}, the Wen plaquette model\cite{wen_plaquette}, as well as recent constructions obtained by decorating lines on the lattice with $1d$ topological superconductors formed by the $\epsilon$ particles\cite{rao2021theory,kane2020}. 

\begin{figure}
 \captionsetup{justification=raggedright}
    \centering
        \adjustbox{trim={.12\width} {.4\height} {.1\width} {.14\height},clip}
    {\includegraphics[width=.7\textwidth]{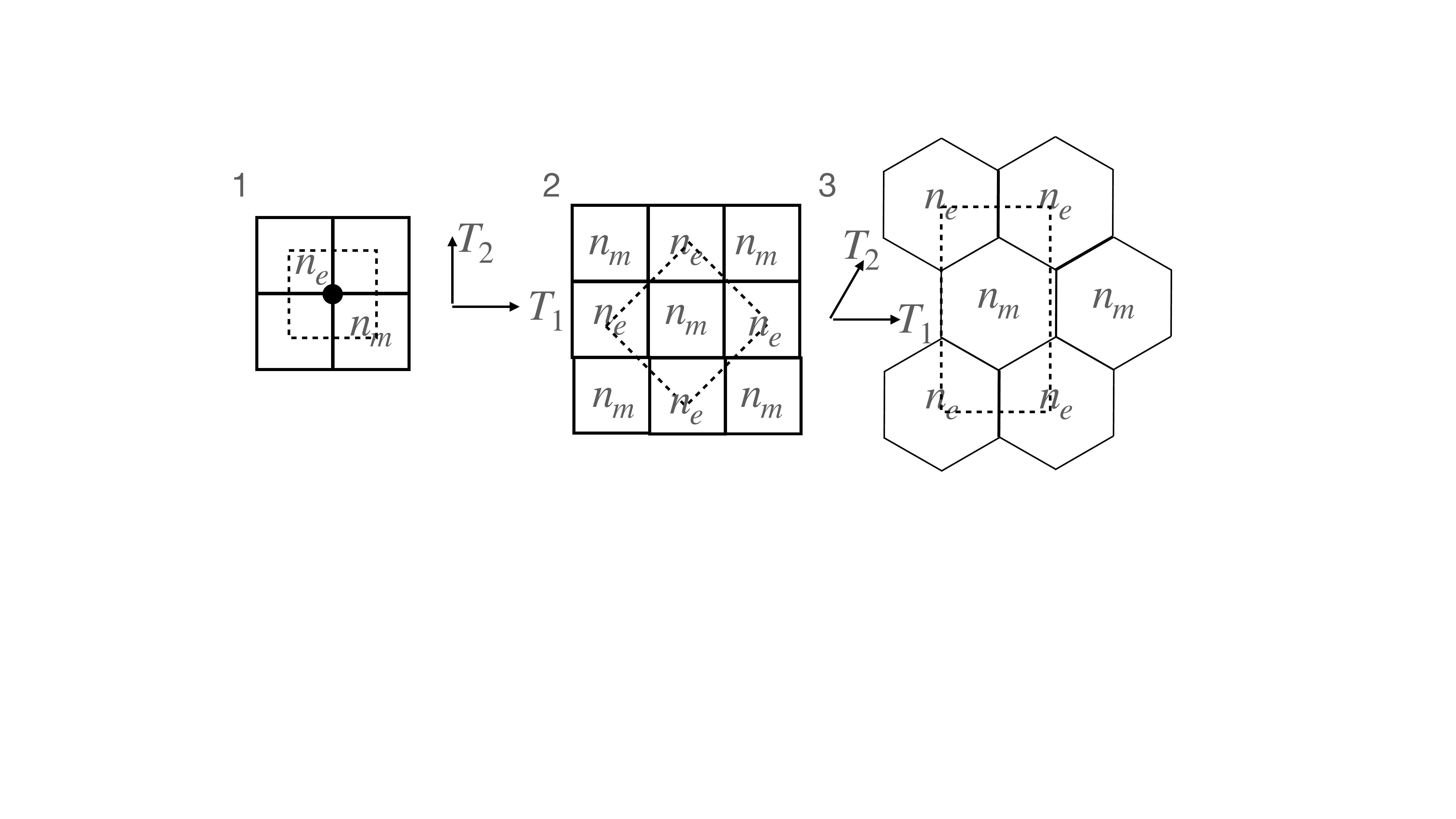}}
    \caption{The model for $Z_2$ QSLs that realize the $12$ SET classes in table \ref{tab:SET}. Model $1,2,3$ are  toric code\cite{kitaev2003}, Wen's plaquette model\cite{wen_plaquette} and the strong bond limit of the Kitaev model, respectively. $n_{e,m}=0,1$ are the number of visons in each model. $\epsilon$ lives on site (or slightly displaced if coincides with $e$). Dashed lines display the trajectory an anyon takes in the defining relation in table \ref{tab:SET}.}
    \label{fig:model}
\end{figure}

\begin{figure}
 \captionsetup{justification=raggedright}
    \centering
        \adjustbox{trim={.022\width} {.001\height} {.04\width} {.08\height},clip}
    {\includegraphics[width=.55\textwidth]{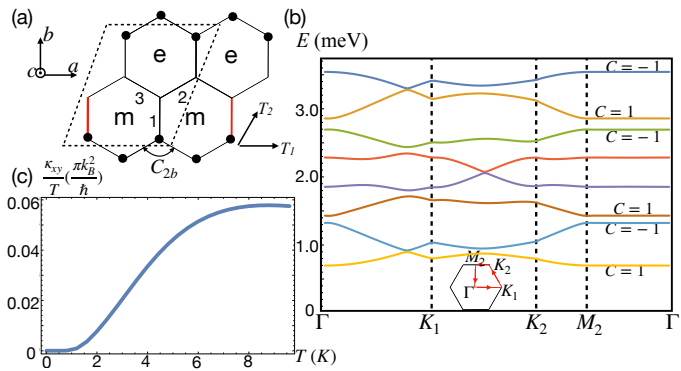}}
    \caption{(a) The honeycomb vison model for $e$ with an $8$-site unit cell and Chern bands,where translation $T_2$ permutes $e,m$.  Visons live on-site and hop in a background anyon flux marked in hexagons. $e,m$ see $\pi(0),0(\pi)$ flux per hexagon in alternate rows along $T_2$, respectively. Hence the hopping and pairing are negative for $e$ on the red bonds. The nearest-neighbor hopping strength is $0.4$meV,  pairing amplitude $0.28 $meV and phase $0,\pm 2\pi/3$ for bond types $1,2,3$, respectively. Chemical potential $-2\pm 0.4$meV for sites with(without) black dot markers, respectively. The model for $m$ is obtained by applying $T_2$ on model for $e$.(b) The dispersion of all $8$ gapped vison bands along high symmetry lines in the \emph{reduced} Brillouin zone. Band touching appears between $4$th and $5$th bands. Chern numbers for other bands are marked. (c) The thermal hall conductivity of $e,m$ in the vison model.The value of $\kappa_{xy}/T$ is comparable to results in ref \cite{Ong2021}.}
    \label{fig:alphamodel}
\end{figure}

In general all physical (local) operators should obey $T_1T_2T_1^{-1}T_2^{-1}=1$.  Restricting first to the simpler situation in which anyons are not permuted under translation (dubbed as type $A$),  the  $e$ (or $m$) could transform with $T_1T_2T_1^{-1}T_2^{-1}=\pm 1$. A $-1$ factor  is the only allowed possibility, such that the physical states with $2$ identical anyons (eq \eqref{emfusion}), still follow a linear representation of translations. Whether the translations act linearly or projectively can be chosen independently for the $e$ and $m$ particles, apparently giving four distinct choices. However keeping in mind that the label $e$ and $m$ can be interchanged without changing the phase, there is a unique phase where translations act projectively on one of $e,m$ and linearly on the other.  Thus we have 3 distinct $Z_2$ QSLs where translations do not permute anyons. To explicitly construct anyon models that realize all  of these possibilities we consider $Z_2$ gauge theories with background $e$ or $m$ charges placed on all the unit cells of the lattice\cite{sachdev,senthil2000} (see Table \ref{tab:SET} and Fig \ref{fig:model}).  Note that the phase labeled $A^\epsilon$ can be understood as having a background $\epsilon$ anyon at each site of the lattice. 

%All physical (local) operators  while the anyons transform in a projective way \cite{wen_psg}(dubbed as projective symmetry group (PSG)), i.e. $T_1T_2T_1^{-1}T_2^{-1}=\pm 1$. The projective representation of physical symmetries on anyons, also called symmetry fractionalization, is the defining relation that is used to classify SETs \cite{zaletel_kagome}.%Here one complication is that translation may permute anyon types, e.g. $e\leftrightarrow m$. In this case $T_1T_2T_1^{-1}T_2^{-1}$ may not be well-defined, then it is pertinent to consider a combined translation that preserves the anyon type, e.g. $T_1T_2, T_1^{-1}T_2$ etc and classify whether such operations commute on the anyons. 

Next we turn to translations that permute $e$ and $m$ particles.  Now either only one of the two translations $T_1$(type $B$) and $T_2$(type $C$) may permute these anyons or both of them may do so(type $D$).  In all cases  we can find a larger translation $T_{\tilde i}$ that preserves the  anyon type.  We then focus on  `elementary' $T_{\tilde i}$'s which generate all translations that act on  a particular anyon without permuting it to a different anyon. For example, for Wen's plaquette model\cite{wen_plaquette} plotted in fig \ref{fig:model} (2) where translations permute $e,m$, the $T_{\tilde i}$'s are taken to be $T_1T_2^{\pm 1}$ indicated by the dashed lines. $T_{\tilde i}$'s are the elementary translations for plaquettes where $e$ or $m$ anyons live, i.e. a checkerboard pattern on the square lattice. %For cases where translations preserve anyons, $T_1T_2T_1^{-1}T_2^{-1}=\pm 1$ suffices to determine the SET in this category, i.e. the classification is exhaustive. 
When translations permute $e,m$, phases are distinguished by whether $T_{\tilde i}$'s  are realized linearly or projectively, as listed in Table \ref{tab:SET}. When different anyon sectors are connected by translations, $e,m$ must transform in the same manner - translations either commute or anti-commute, denoted by a superscript $\pm$ attached to the translation types $B,C,D$. 

Finally let us consider phases obtained by placing a background $\epsilon$ anyon at each site. 
 We argue that for $B^+,C^+,D^+$ types, this generates new phases while for $B^-, C^-,D^-$ it does not.  Translation $T_i$ permuting $e,m$ can be realized by placing a series of Kitaev chains\cite{kitaev_2001} along the transverse direction, 
which induces a Majorana when a $\pi$ flux ($e,m$) tunnels across the wire, and exchanges $e,m$\cite{rao2021theory,kane2020}.  Now for the $+$ phases, consider the allowed ground states on a torus with $L_xL_y$ unit cells  ($L_{x,y}$ the number of sites along $x$ and $y$ directions respectively), and add a background $\epsilon$ on each site: to conserve fermion parity per the gauge constraint, the  boundary conditions (BC) for the Kitaev wires  should be changed so that there is an extra $\pi$ flux compared to when there was no background added $\epsilon$ charge. This change in the allowed ground states on such a torus is probed by asking about the phase seen by the $\epsilon$ as it is dragged around the appropriate cycle of the torus.  Hence we conclude that these are new SET classes.   Stacking $\epsilon$ per site, however, does not change the defining relations for $e,m$ of modified $T_{\tilde i}$, as the trajectory for the relations enclose an even number of $\epsilon$'s.  For SET $B^-,C^-,D^-$ where background $e,m$ exists, the additional $\epsilon$ can be combined with background visons and does not give new SET classes\footnote{Here the criterion of BC change does not apply since a consistent assignment of background $e,m$ to each plaquette, requires even dimension along the direction that permutes $e,m$.}. We name these classes with background $\epsilon$ on each site with a superscript $\epsilon$ in table \ref{tab:SET}. 

We thus have $12$ translation-enriched gapped $Z_2$ QSLs in $2$d, listed in table \ref{tab:SET} . To construct models for each SET class we use the $3$ prototypes in fig \ref{fig:model} to realize different translation types with details in 
Appendix \ref{appset}.

In passing, we note that we can readily generalize the classification to $3+1D$ $Z_2$ topological order  with point-like bosonic $e$ particles and tensionful unoriented vison loops as fractional excitations. We show in Appendix \ref{app:3dset}\cite{levingu,wang_weyl} that there are $128$ symmetry enriched $Z_2$ QSLs in $3D$, distinguished by defining relations on $e$, vison loops and decorating lattice planes with $2d$ symmetry protected topological states of the bosonic $e$-particles.

\begin{table}
\captionsetup{justification=raggedright}
\centering
\begin{tabular}{|c|c|c|c|c|}
\hline
 Type & $T^\epsilon_{12}$& \multicolumn{2}{|c|}{Defining relations on $e,m$}& Model\\ \hline
$A^+$ & $1$ &&$T^{e(m)}_{12}=1$ & 1\\
$A^-$ &$-1$ & &$T^{e(m)}_{12}=\pm 1$&1: $n_{e(m)}=1(0)$\\ 
$A^\epsilon$ &$1$ &not for odd $L_xL_y$& $T^{e(m)}_{12}=-1$ &1: $n_{\epsilon}=1$ \\ \hline
$B^+$ &$1$&\multirow{2}{*}{$T_{\tilde 1}=T_1^{2}$} &$T^{e(m)}_{\tilde 12}=1$ & 3\\\ 
$B^-$&$-1$ && $T^{e(m)}_{\tilde 12}=-1$ & 3: $n_{e(m)}=1$ \\ \
$B^\epsilon$&$1$&BC change& $T^{e(m)}_{\tilde 12}=1$ &3:  $n_\epsilon=1$\\ \hline
$C^+$ &$1$&\multirow{2}{*}{$T_{\tilde 2}=T_2^{2}$} &$T^{e(m)}_{1\tilde 2}=1$ & 3\\\ 
$C^-$&$-1$ && $T^{e(m)}_{ 1\tilde 2}=-1$ & 3: $n_{e(m)}=1$ \\ \
$C^\epsilon$&$1$&BC change& $T^{e(m)}_{ 1\tilde 2}=1$ &3: $n_\epsilon=1$\\ \hline
$D^+$ &$1$&\multirow{2}{*}{$T_{\tilde 1,\tilde 2}=T_1^{\pm 1}T_2$} &$T^{e(m)}_{\tilde 1\tilde 2}=1$ & 2\\\ 
$D^-$&$-1$ && $T^{e(m)}_{\tilde 1\tilde 2}=-1$ & 2: $n_{e(m)}=1$ \\ \
$D^\epsilon$&$1$&BC change& $T^{e(m)}_{\tilde 1\tilde 2}=1$ &2: $n_\epsilon=1$\\ \hline
\end{tabular}
\caption{Classification of $Z_2$ QSLs with translations.  Naming rules: (A) translations preserve anyon types, (B) only $T_1$,and (C) only $T_{2}$ permutes $e,m$ and (D) $T_{1,2}$ both permute $e,m$. Superscript $\pm,\epsilon$ denotes no background anyon,   background $m,e$ or a background $\epsilon$ per site, respectively. %\sout{An $\epsilon$ atomic insulator would change the boundary condition (BC) on a torus with an odd number of sites $L_xL_y$.}  
Defining relations $T^{\epsilon,e,m}_{ij}=T_iT_jT_i^{-1}T_j^{-1}$ for anyons are listed. Models $1,2,3$ are shown in fig. \ref{fig:model}. Translation types $B$ are realized by rotating model $3$ by $120^o$ clockwise. $n_{e,m,\epsilon}$ are the number of $e,m,\epsilon$ per site/plaquette which has default value $0$ unless specified in the table.%\sout{SET with superscripts $+,\epsilon$ forbid a Chern band on primitive Bravais lattices. Square, honeycomb and kagome FFIM are realized by SET $A^-$, while triangular FFIM by SET $A^+$.
\label{tab:SET}}
\end{table}

Armed with this understanding we turn next to the question of which of these $2d$ $Z_2$ QSLs admits vison Chern bands. A quadratic Hamiltonian provides an adequate description of the gapped visons, so that the notion of a Chern number is well defined for the resulting bands. 
We prove in Appendix \ref{app:proof}\cite{rmp_berry,magnon_shindou,magnon_rev} that since translation-invariant SETs where translations commute on anyons, are governed by an explicitly translation-invariant Hamiltonian, no vison Chern band is possible if the visons hop on a primitive Bravais lattice. %\sout{i.e. contains one site per unit cell. This does not exclude a nonzero thermal hall conductivity since $\kappa_{xy}$ is a weighted sum of the Berry curvature, though it could guide the search for topological vison bands which more naturally contribute to thermal hall effects.}

%\sout{Here instead we list general principles for constructing vison models given minimal information on a system: First determine the symmetry of the system. In the paper we are concerned with translations only, as in the measurement of $\alpha$-RuCl$_3$ under a magnetic field in a generic orientation, rotations, time-reversal, reflection or their combination are broken, leaving only translation intact. Next find out the mechanism for fractionalization and $Z_2$ gauge structure, e.g. from parton construction. This helps (partially ) determine how translations act on visons and the SET class.} % For example for a conventional slave-particle decomposition of spin$-1/2$ (e.g. fermionic spinon, Schwinger boson etc) lattices, there is one slave particle per site and the visons living on dual lattices see a $\pi$ flux around the slave particle. Hence we constructed the fully-frustrated Ising paramagnet model below. 

The presence of background anyons, if any,  leads to a braiding phase seen by other anyon species on any path that encloses an odd number of the background anyons,  with  discussions for generic $Z_2$ QSLs in Ref \cite{chen2022}. Specifically if we consider, eg, hopping of an $e$ particle, 
\begin{eqnarray}
\label{fluxcons}
\textrm{Sign}[\prod_{\bar{ij}\in p} t_{ij}]=(-1)^{n_p}(-1)^{l_p},
\end{eqnarray} 
where $p$ is any path formed by vison hopping $t_{ij}$, $l_p$ the number of hops along $p$, and $n_p$ the number of background anyons ($m$ or $\epsilon$) enclosed by $p$. 
%Further neighbor hopping are allowed under the same flux constraints. This implies that hopping flux should be translation-invariant or hopping flux for $e,m$ should relate to each other by translation if they permute $e,m$.%\footnote{Pairing could be added which is translation-invariant  under the projective translations determined by NN hopping.} 
%Strictly speaking, for trajectories that are fractions of an elementary plaquette, the flux should be $0, \pi$ is the trajectories encircle $0$ or odd number of gauge charges, where gauge charges are localized at a point in space, e.g. the plaquette center.

\emph{Topological bands from fully frustrated visons- }A common spin liquid state constructed by parton constructions for spin-$1/2$ operators (Schwinger boson or Abrikosov fermion) has one `spinon' per site, conveniently viewed as a background anyon of some fixed type.  Without loss of generality we take this background anyon to be either an $e$ or $\epsilon$ particle. In such an SET state the visons  ($m$ particles) are described by a fully frustrated Ising model(FFIM) on the dual lattice\cite{sachdev,senthil2000,moessner,moessner2}.  The frustration is a direct consequence of the braiding phase of $m$ around the background anyon. % implies that the product of nearest-neightbor hopping amplitudes $t_{ij}$ around \emph{elementary} plaquettes $p$ differs in sign from those of a ferromagnetic Ising couplings for visons, i.e. $n_p=1$ in eq \eqref{fluxcons}. This class of model has also been studied in ref \cite{sachdev,moessner,moessner2} in the context of doping 2D frustrated antiferromagnets and dubbed as fully frustrated Ising paramagnets (FFIM).

We construct (projective) translation invariant models for visons on $3$ kinds of lattices which exhibit Chern bands. The Hamiltonian for the visons formally reads,
\begin{align}
\label{eq:ffim}
    H_{v}=\sum t_{ij} v_i^\dagger v_j+\sum \Delta_{ij} v_i^\dagger v_j^\dagger+h.c.,
\end{align}
where $v_i^\dagger$ is a vison creation operator at site $i$. Pairing is added since vison numbers are conserved modulo $2$. 

On dual square, honeycomb lattices, the hopping flux around an elementary plaquette is $\pi$ for visons to be frustrated, while on triangular lattices the flux is $0$ and on kagome lattices flux is $0,\pi$ for triangle/hexagon plaquettes, respectively. Hence the translations for visons anticommute on square, honeycomb and kagome lattices. Translations, however, commute on triangular lattices. This shows that it is \emph{impossible} to have a Chern band on triangular FFIM. We list the model with topological vison bands and the typical Berry curvature distribution in appendix \ref{app:ffim}. This exercise demonstrates (not surprisingly) that band topology of visons is possible as a matter of principle in $Z_2$ QSLs. Whether it is actually realized is a question to be answered by calculations on microscopic models or through experiment.  

% We explore the possible origin of the thermal hall conductivity in ref\cite{Ong2021}.
 %Besides the spin-wave theory\cite{magnon_ybkim,magnon_moessner,magnon_joshi,moore18}, . %For exotic possibilities, under the Schwinger boson formulation for spin liquids, with only translations preserved, the energy minimum that the spinons condense into is likely to be at a generic momentum rather than $\Gamma$ point, resulting in a non-collinear magnetic order \cite{sachdev_bspinon,wang_bspinon}. \footnote{Since spin rotation is completely broken, one could not assign spin quantum numbers and there is no definite distinction between spinons and visons.} 
% We resort to the flux excitation- visons - in a $Z_2$ spin liquid state. %(arising from different origins than Schwinger boson formulations). 
\emph{Application to $\alpha$-RuCl$_3$:} Microscopic calculations on realistic models (significantly different from the pristine Kitaev model) of $\alpha$-RuCl$_3$ are difficult and beyond the scope of this work. We will instead develop a phenomenological model for the vison bands that, as will see, accommodates some of the key experimental features of this material\footnote{Some thermal Hall measurements in  an out-of-plane $B$\cite{alpha_phonon} are proposed to be due to phonons. Theoretically the relevance of phonons to thermal hall in  diverse systems have been investigated \cite{ye_thermal,rosch_phonon,chen_phonon,guo_phonon,phonon_nonabelian}; here, the phonon origin is inconsistent with the $\kappa_{xy}$ scale and temperature dependence in $\alpha$-RuCl$_3$\cite{Ong2021}}. 
Specifically we introduce a vison model on the honeycomb lattice where one translation $T_2$ permutes $e,m$. To have nontrivial vison band topology, we consider background anyon flux as a source of projective translation for the visons and assign $e,m$ to live in hexagons in alternate rows along $T_2$.  This falls into SET class $C^-$.   Besides hopping and pairing, a staggered chemical potential is added to reflect the weak translation breaking for visons and dynamically suppress nearest neighbor hopping of $e$ or $m$ along $T_2$ which excites an $\epsilon$. Fig \ref{fig:alphamodel} shows the model with translations and $C_{2b}\mathcal T$, band structure and $\kappa_{xy}/T$ (details in Appendix \ref{appformula}\cite{murakami}).  

There are $3$ advantages to this construction: 1)  A key feature of the putative intermediate field QSL is its proximity to the zigzag magnetically ordered phase at low $B$, which breaks one translation $T_2$ and preserves $T_1$(also $C_{2b}\mathcal T$ if ordered in $a-c$ plane). This is the only distinction from the QSL phase symmetry-wise. Since condensing $e$ or $m$ would break $T_2$, this model naturally accounts for the zigzag phase at small $B$ that preserves $T_1,T_2^2,C_{2b}\mathcal T$  provided visons condense at the $\Gamma$ point, which may be natural for small $B$. 2) The  topological vison band structure allows a thermal Hall effect that will  decrease to zero as $T \rightarrow 0$.  
The change of sign in $\kappa_{xy}$ when $B$ flips direction in the $ab$ plane is guaranteed since this corresponds to the time-reversal breaking terms for visons to change sign, which changes the sign of the Berry curvature and hence $\kappa_{xy}$. This route to the thermal Hall signal is similar to that in the topological magnon theory, which presumably applies at higher fields beyond the proposed QSL regime). Heuristically as $B$ is decreased the magnons fractionalize into visons upon entering the QSL; it is perhaps not surprising that the resulting vison bands inherit a band topology from the parent magnons. On the technical side, despite the resemblance to the theory of topological magnon bands,  the presence of background anyon flux allows frustrated hopping for $e,m$, and hence offers more possibility for  development of Chern bands for visons. 3) The fractionalization of magnons into visons implies that the neutron scattering signal will be dominated by continuum scattering and will not see sharp peaks. This is indeed what is seen in neutron experiments\footnote{In contrast ESR \cite{esr} sees sharp resonances even in the putative QSL region, at frequencies accessible to neutron spectroscopy. It is unclear at present how to reconcile these two observations. }    

A striking experimental signature of the  translation-enrichment in this QSL is an enhanced periodicity of the spin structure factor associated with the continuum scattering within the microscopic Brillouin zone. This was discussed in Ref. \cite{hermele_neutron} and readily generalized to situations where translation permutes the anyons. For the specific $Z_2$ QSL above, the enhanced momenta are discussed in appendix \ref{app:neutron}.   Existing experiments show weak signals at these momenta. Our  discussion urges further experiments to look for the enhanced periodicity. %The  general ideas explored here will prove broadly useful for putative gapped QSL systems.   

\emph{Considerations from microscopics} A $Z_2$ QSL where one translation permutes anyons occurs already in a spatially anisotropic deformation of the Kitaev model. This model has no background anyon ({\emph i.e} zero flux) in the ground state. Realistic models for $\alpha-RuCl_3$ include a significant additional \emph{antiferromagnetic} $\Gamma$ interaction\cite{plumb_gamma} between nearest neighbors,
$\sum_{\langle ij\rangle_\gamma}\Gamma S_i^\alpha S_j^\beta,$
where $\alpha,\beta$ are the two remaining spin indices for a $\gamma$ type bond in Kitaev model. The inclusion of the $\Gamma$ terms  stabilizes an apparently distinct QSL dubbed $K\Gamma SL$ in Ref \cite{kee_nc}, with a negative flux expectation value ($\langle W_p\rangle$). If this is a $Z_2$ QSL, this signals a state with background vison numbers $n_{e,m}$ per plaquette. The phase diagram in a field is highly sensitive to small perturbations (anisotropic interactions and Heisenberg exchange). It is conceivable that for some choice of parameters a $Z_2$ QSL may emerge at intermediate fields, distinct from that associated with the exactly solvable Kitaev model (as the existence of $K\Gamma SL$ already demonstrates). Our general discussion of translation enriched topological order helps identify such a QSL in future studies of microscopic models. % Additional further neighbor $\Gamma'$ interactions drives the system at small $B$ into zigzag and favors negative $W_p$ as well. At 3rd order perturbation, $\Gamma$ interaction along $3$ non-adjacent bonds in a hexagon induces $\frac{\Gamma^3}{K^2} W_p$ which favors nonzero $n_{e,m}$. For in-plane $B$ this prompts an intermediate $K\Gamma SL$ regime (between zigzag and polarized states) with nonzero $n_{e,m}$, as we assumed in the model in fig \ref{fig:alphamodel}. }

\emph{Conclusion} Motivated by indications for an intermediate-field QSL in $\alpha$-RuCl$_3$, we studied some general properties of gapped $Z_2$ QSLs with no global symmetries other than lattice translations. Building on prior partial results we classified such QSLs, and identified situations in which visons ($e$ and $m$ particles) form topological Chern bands. We discussed some phenomenologically attractive features of such gapped $Z_2$ QSLs with vison Chern bands: continuum scattering in neutron spectroscopy, a thermal Hall effect that decreases to zero with temperature, and a natural connection to zigzag magnetic ordering. The models discussed naturally have images of the continuum scattering at multiple locations in the Brillouin zone. Our discussion is phenomenological, and making contact with microscopic calculations will be important moving forward.  On the theoretical front, the critical confinement(Higgs) transition from the specific $Z_2$ QSL discussed to the zigzag phase would be interesting to study numerically. Closely related transitions have been studied numerically\cite{nahum2020} with fascinating results that are yet to find a field theoretic understanding.

\emph{Acknowledgements}: We thank Stephen Nagler, Hae-Young Kee and Subir Sachdev for useful discussions. We also thank Hae-Young Kee for a careful reading and comments on an earlier version of the paper. X-Y Song thanks discussion with Yi-Zhuang You, Tarun Grover and John McGreevy. X-Y S is supported by the Gordon and Betty Moore Foundation EPiQS Initiative through Grant No. GBMF8684 at  MIT. This work was supported by NSF grant DMR-1911666,and partially through a Simons Investigator Award from the Simons Foundation to Senthil Todadri. This work was also partly supported by the Simons Collaboration on Ultra-Quantum Matter, which is a grant from the Simons Foundation (651440, TS).  

\appendix
\section{Symmetry enrichment of $Z_2$ topological order with translations}
\label{appset}

%The $n_{e,m}=0,1$ marked the location of anyon excitations and their number in the ground state. 

Here we list the model construction plotted in fig 1 of the main text for the realizations of different translation-enriched SETs.

To realize the defining relation i.e. projective symmetry group (PSG) in each SET class, we utilize the fact that $e,m$ have mutual $\pi$ braiding statistics. An adiabatic evolution around the other type of anyon could be realized by the translation operation $T_1T_2T_1^{-1}T_2^{-1}$. Hence the phase factor $\pm 1$ in the relation depends if the trajectory of the translation operation encircles an odd number of anyons of the other species. The anyon occupation can be tuned by adjusting the chemical potential for the anyons. 

For type $A^{\pm,\epsilon}$, we look to the original Kitaev toric code model where $Z_2$ gauge field $\sigma^z$ lives on the square lattice bonds\cite{kitaev_2006}.
\begin{align}
H_{tc}=\mu_m\sum\prod_{i\in P} \sigma^z_i+\mu_e\sum\prod_{i\in v} \sigma^x_i,
\end{align}
where $P,v$ denote the elementary plaquette and the vertices, respectively. $\mu_{m,e}$ are the chemical potential for an $m,e$ excitation, respectively. When $\mu_{m,e}>0$, the ground state is the vacuum and the translations on the visons permute. When $\mu_m\mu_e<0$, the ground state contains one $e$ or $m$ particle per unit cell depending on which chemical potential is negative. Hence the translations for $m$ or $e$ anti-commute due to the mutual semion statistics of the visons. Note this gives one unique SET class $A^-$ since one can always define the vison that has anti-commuting translations to be e.g $e$ particles.

When $\mu_m<0,\mu_e<0$, there is one $e$ and $m$ (identically an $\epsilon$) per unit cell, and translations for both anyons anti-commute, hence type $A^\epsilon$.

For the model where one translation permutes $e,m$, we look to a variant of the toric code model as discussed in sec 7.1 of ref \cite{kitaev_2006}. Plotted in fig 1(3) of the main text, the $e,m$ particles live in alternate rows of the honeycomb lattice. One again tunes the chemical potential. When $\mu_{e,m}>0$, the ground state has no anyon charge and translations for anyons commute. When $\mu_m=\mu_e<0$,there is one $e$ and $m$ per unit cell, and translations for both anyons anti-commute, hence class $C^-$. Note the chemical potential for $e,m$ should be equal due to the translation that permutes the anyons. One may ask if the projective relation could be different for $e,m$. It is indeed possible when neither translation permutes anyons that $e,m$ obey different projective translations. Hence type (a) further classifies into $3$ cases where both/neither anyon  transform projectively (i.e. two translations commute), or one of $2$ anyons transform projectively. 
Type $B^{\pm}$ can be similarly obtained by rotating the model in fig 1(3)(main text) $120^o$. Such a $Z_2$ topological order can be constructed on the square lattice, e.g. by coupling together 1d Kitaev  chains and gauging the fermion parity\cite{rao2021theory,kane2020}. 

When both translations permute $e,m$, we take  Wen's plaquette model\cite{wen_plaquette}. The $Z_2$ variables live on the sites and the Hamiltonian reads,
\begin{align}
H=\mu \sum\prod_{i\in A} \sigma^z_i+\sum\prod_{i\in B} \sigma^x_i,
\end{align}
where $A,B$ denote the elementary plaquettes that organize into a checkerboard pattern on square lattice as plotted in fig 1 (2) in the main text. $\mu>0$ yields a trivial ground state and class $D^+$. $\mu<0$ gives one vison per plaquette as the ground state and translations anti-commute, hence class $D^-$. 

For classes with superscript $\epsilon$, they can be readily realized by the corresponding model (toric code for $A$, model 3 for $B,C$ and Wen's plaquette model for $D$, respectively) by placing one $\epsilon$ per site.

\section{Thermal hall from a vison Chern band}
\label{appformula}

The thermal hall conductivities from bosonic particles arise from integral of the Berry curvature\cite{rmp_berry} multiplied by energy-dependent functions\cite{murakami}. Hence generally a band of nontrivial Chern number will give notable thermal hall conductivity.

\begin{align}
\kappa_{xy}=\frac{2 k_B^2T}{\hbar V}\sum_{n,\vec k} c_2(\rho_n)Im \langle\frac{\partial u_n}{\partial k_x}|\frac{\partial u_n}{\partial k_y}\rangle.
\end{align},
where $\rho_n\equiv \rho(\epsilon_{nk})$ and $c_2(\rho)=(1+\rho)(\log \frac{1+\rho}{\rho})^2-(\log \rho)^2-2Li_2(-\rho)$, $Li_2(z)$ is the polylogarithm function.

 When calculating the thermal hall conductivity for the model we propose in the context of $\alpha-$RuCl$_3$, the different bands are not completely isolated from each other. %Consider pure boson hopping model, there is no difference in the particle statistics, i.e. being a boson or fermion. 
The band touching points are described by (single particle) Dirac Hamiltonians and the thermal hall contribution of the corresponding states   vanishes. These states have the same energy and the Berry curvature adds to $0$ in proximity of the Dirac point.

\section{Translation-commuting vison models have trivial band topology}
\label{app:proof}
We prove that in the SET classes where translations commute, i.e. class $A^{+,\epsilon},B^{+,\epsilon},C^{+,\epsilon},D^{+,\epsilon}$ in Table I of the main text, it is impossible to generate Chern bands when the visons hop on a unit cell with just one site. For SETs where translations permute anyons,  we consider effectively only the part of the lattice where a particular anyon lives and a model on the effective lattice, since $e,m$ cannot mix as they live in different superselection sectors. Then the composite translation $T_{\tilde i}$ that we use in Table I of the main text for classes IV, VI,VIII act as elementary translations on the effective lattice.

First in a projective relation of translations $T_{1,2}$ where the gauge group  is $Z_2$, the phase for $T_1T_2T_1^{-1}T_2^{-1}$ is equivalent to the hopping flux along the trajectory $p$ of a certain site acted on by $T_1T_2T_1^{-1}T_2^{-1}$, ie.
\begin{align}
\label{cl1}
Arg[T_1T_2T_1^{-1}T_2^{-1}]=Arg[\prod_{l\in p} t_l].
\end{align}
 Note that the hopping $t_l$ is oriented along the trajectory of the site. Consider a projective translation by a vector $\vec v$,
\begin{align}
T_v: c_i\rightarrow e^{ig_v(i)}c_{i+\vec v},
\end{align} 
where a phase is encoded in the function $g_v$. Then for a site $i$
\begin{align}
\label{projectivephase}
Arg[T_1T_2T_1^{-1}T_2^{-1}]=(g_1(i)-g_1(i+\vec l_2))+(g_2(i+\vec l_2)-g_2(i)).
\end{align}

One further knows that for the translation to keep the hopping invariant i.e.
\begin{align}
T_v: c_i^\dagger c_j\rightarrow e^{i (g_v(i)-g_v(j))} c_{i+\vec v}^\dagger c_{j+\vec v},
\end{align}
 the phase difference satisfies
\begin{align}
\label{diff}
g_v(i)-g_v(j)=Arg\left[\frac{\prod_{l\in i+\vec v\rightarrow j+\vec v}t_{l}}{\prod_{l\in i\rightarrow j}t_{l}}\right],
\end{align}
where the hopping amplitude is multiplied along a (series of) bonds $l$ that connect $i,j$.

Combining eqs \eqref{projectivephase} and \eqref{diff}, we prove eq \eqref{cl1}.

For the SET classes where translations commute, it is always possible to choose a gauge such that all $t_l$ around the trajectory $p$ is real and positive, so that the flux to be $0$. In this case, the translation action is trivial, i.e. $g_v=0$ identically. It is still true even with vison pairing or hopping beyong nearest-neighbor.  

Hence the vison model is explicitly translation invariant. In this case, if the unit cell on which the vison moves contains only one site, the reciprocal space Hamiltonian is factorized as
\begin{align}
H(\vec k)\equiv e^{-i\vec k\cdot \vec R} H(\vec r)e^{i\vec k\cdot \vec R}=h(\vec k),
\end{align}
where $\vec R$ is the Bravais lattice location that labels unit cells. This is a single band system and the Berry curvature /Chern number vanishes. This still holds if one adds pairing. Suppose the Hamiltonian reads in momentum space (with the spinors $\psi_k=(c_k,c_{-k}^\dagger)^T$),
\begin{align}
H(\vec k)=\psi_k^\dagger h_k \psi_k
\end{align}
where $\sigma^1 h_k^* \sigma^1=h_{-k}$ is a $2\times 2$ Hamiltonian. The two bands are split into a particle and hole band, respectively \cite{magnon_shindou,magnon_ybkim,magnon_rev} where each band contains the full information of the system due to the redundancy relating $h_k, h_{-k}^*$. The Chern number sum for the particle/hole band is $0$\cite{magnon_shindou}, which in our case means the single band holds a trivial Chern number.% i.e. a Galilean-invariant system has vanishing thermal hall in the $k\rightarrow 0$ limit \cite{tsc_thermal}. %and $M$ is a constant matrix that has the dimension of (twice) the unit cell size which encodes the hopping (and pairing) structure of the translation invariant Hamiltonian. The eigen function for a given band hence is independent of momentum $\vec k$ and the Berry curvature vanishes. 
 Hence we have proved that no Chern band is possible for translation commuting  SETs. This does not exclude a nonzero thermal hall conductivities since $\kappa_{xy}$ is a weighted sum of the Berry curvature, though it could guide the search for topological vison bands which more naturally contribute to thermal hall effects.

\section{Models with topological vison bands in fully frustrated Ising paramagnets on honeycomb, kagome and square lattices}
\label{app:ffim}

For simplicity, for FFIM we mainly consider vison hopping (including small pairing will not change band topology).  

The symmetries of interests are translations only, which constrains that the hopping flux for visons should be translation invariant. 
 We specialize to  the case where the visons live on the plaquettes of the original lattice which is natural in the parton construction.  Furthermore the parton construction for spin-$1/2$ systems gives natural access to phases where the vison hopping  are frustrated as described in eq (3) of the main text. We construct  a hopping model to demonstrate the possibility of Chern bands.

The hopping amplitudes $t_{ij}$ are plotted in fig 3 by arrows (imaginary hopping) and colors of the bonds (real hoppings).The model is complemented by imaginary hoppings that breaks time reversal, necessary for nontrivial band topology. We plot the Berry curvature distribution for one of the Chern bands on Honeycomb and Kagome lattices in fig \ref{fig:bc}.On square lattices, the model is the familiar spinon mean-field ansatz for a chiral spin liquid that holds a Chern band.

\begin{figure}
 \captionsetup{justification=raggedright}
    \centering
        \adjustbox{trim={.08\width} {.14\height} {.4\width} {.1\height},clip}
    {\includegraphics[width=.8\textwidth]{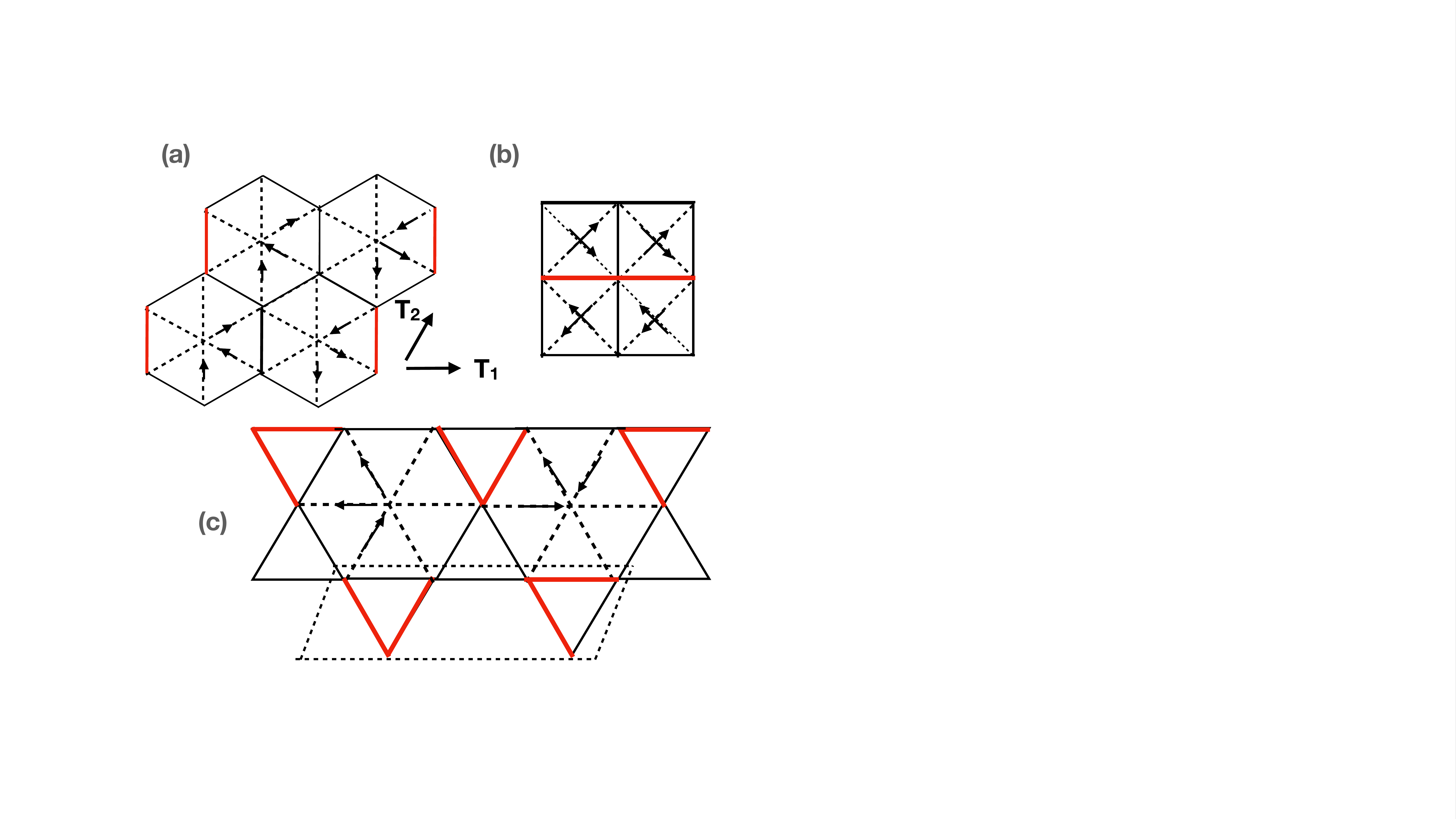}}
    \caption{The fully frustrated vison hopping model that yields a Chern band on honeycomb, square and kagome lattices. The model is translation invariant as the hopping flux around any kind of plaquette is translation invariant. The arrows from $i\rightarrow j$ indicate imaginary hopping $ic_i^\dagger c_j$ and solid black/red lines indicate real positive/negative hopping, respectively. (a) On honeycomb lattice  we take the next-nearest-neighbor(NNN) imaginary hopping to be large that effectively realize a chiral spin liquid on triangular lattices. The nearest-neighbor(NN) hopping satisfy the fully frustration condition. (b) On square lattice the FFIM hop in $\pi$ flux and the diagonal imaginary hopping yields a Chern band. (c) For Kagome lattices the NN hopping is real and diagonal hopping imaginary indicated by the arrows.}
    \label{fig:chernmodel}
\end{figure}

%\section{Chern band construction for FFIM on square, honeycomb and Kagome lattices}
\begin{figure}
 \captionsetup{justification=raggedright}
    \centering
        \adjustbox{trim={.08\width} {.1\height} {.05\width} {.1\height},clip}
    {\includegraphics[width=.6\textwidth]{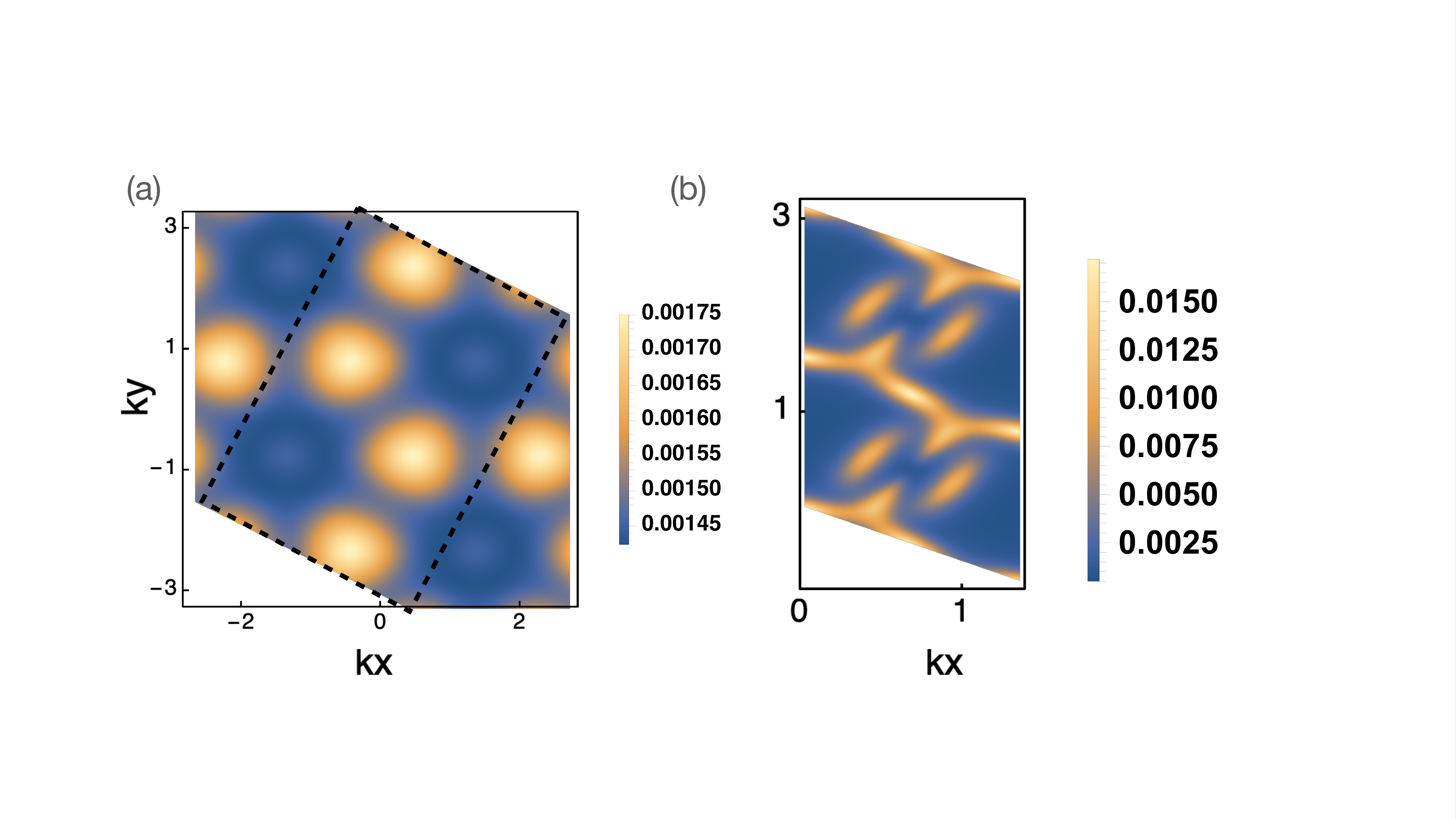}}
    \caption{Berry curvature for the honeycomb (a) and kagome  (b) vison model (FFIM) plotted in fig \cite{fig:chernmodel} which give a Chern number of $1,3$, respectively. Dashed square in (a) and the region plotted in (b) is the reduced Brillouin zone as the unit cell is enlarged by a factor of $2$ for the FFIM.There is a $C_6$ rotation for the FFIM on honeycomb lattice we consider.}
    \label{fig:bc}
\end{figure}

\section{Translation SET classification for $(3+1)$D $Z_2$ topological orders}
\label{app:3dset}

In $3+1$D $Z_2$ topological orders there are gauge charges $e$ and vison loops  $m$. Clearly translations cannot permute particles and loops. We limit the discussion to cases where the gauge charge $e$ is a boson. 

Take a cubic lattice with translations along $3$ orthogonal directions $T_{1,2,3}$ for example where $e$ lives on the sites. There are $3$ projective translation relations for $e$, $T_iT_jT_i^{-1}T_j^{-1}=\pm 1$ where $T_{i,j}(i,j\in[1,2,3])$ are any $2$ orthogonal translations of the $3$ elementary translations. The phase factor depends on the $Z_2$ gauge flux per plaquette in the plane   defined  by $T_{i,j}$. In total there are $2^3=8$ classes from translations on $e$.   Similarly new SET classes are obtained by asking if a gauge charge is placed at each site. Having $1$ gauge charge per site in the background will manifest in the translation of vortex loops: Consider a vortex loop along e.g. the $T_1$ direction in a system with odd dimension $L_1$ along $T_1$, and move it by $T_2T_3T_2^{-1}T_3^{-1}$: it will then accumulate an $L_1\pi ~mod~ 2\pi=\pi$ Berry phase from encircling the gauge charges on $L_1$ sites. %The defining relation for vortex loop is determined by a single relation $T_iT_jT_i^{-1}T_j^{-1}=\pm 1$ for any pair of $(i,j)$. Hence there are $2$ possibilities for vortex loop translations.  }
%For translations of the vison loop $m$, for concreteness we consider a loop in the $1-2$ plane with dimension $1$ along $T_1$ {\color{red}  The relation $T_2T_3T_2^{-1}T_3^{-1}=\pm 1$ for one end of such a vortex loop is given by number of gauge charges on the site encircled by such an operation, as one can explicitly write down the operators for translations of vortex loops and see that it is the gauge charge on the site. Since the model is translation invariant, one can have $0$ or $1$ unit of gauge charge per site identically. 
Hence we can view the Berry phase $0,\pi$ of vortex loop translation under this particular setting as corresponding to decorating each lattice site with either none or a single $e$ particle, {\em i.e}, to placing $0+1$-dimensional Symmetry Protected Topological (SPT) states of the $e$ particles on lattice sites. In total there are $16$ classes for 3D $Z_2$ translation SET orders from placing background gauge charges or flux loops.

Following ideas similar to our discussion in $2+1$-d, further distinct phases can be obtained by decorating lines with $1+1$-d SPTs, or planes with $2+1$-d SPTs of the $e$ particles. However there are no $1+1$-d SPT states protected by $Z_2$ symmetry. In $2+1$-d there is a well-known boson SPT state with $Z_2$ symmetry which is nicely illustrated by the Levin-Gu model\cite{levingu}.
%{\color{blue}In the case for $2+1$d classification,placing an $\epsilon$ on each site leads to new  SET classes,i.e. stack lower-dimensional symmetry-protected topological(SPT) states and gauge the $Z_2$ symmetry in order to construct new topological orders enriched by symmetries. Here one can similarly ask about stacking $2+1$d $Z_2$ topological orders of bosonic systems. One candidate is the Levin-Gu model \cite{levingu} with a $Z_2$ Ising-like symmetry. 

Gauging the Levin-Gu model leads to the so-called double-semion state, where a $\pi$ flux is a semion with $\pi/2$ self statistics and when binding with a gauge charge becomes an anti-semion. Stacking these SPTs of the $e$-particles, in e.g. $1-2$ plane along the $T_3$ direction, will result in a semion excitation at the core of a vison loop every time it crosses a $1-2$ plane. This can be probed by considering two vison loops linked with a dislocation half-plane ($1-2$ plane) with a Burger's vector of a unit vector along $T_3$. In the $2$-loop braiding process linked with a dislocation plane\cite{wang_weyl}, where one loop first shrinks and goes through the other loop, and then enlarges and crosses back (encompasses) the other loop again, the $2$ semions trapped at the half-plane go through a $2\pi$ braiding process resulting in a $\pi$ berry phase coming from their statistics. Next consider combining such a  stack of  e-SPTs  on planes with background $\pi$ flux through each plaquette. Then   a semion will be bound on each plaquette. Again a vison loop linked with a dislocation half-plane (where the e-SPT lives) will see a $\pi$ Berry phase when moving around a plaquette with a background semion. Similarly combining stacking the e-SPT on planes with states with background  e-particles on sites will make the $3$-loop braiding when linked with a dislocation half-plane dependent on how many sites with gauge charges are encircled and reflect the additional contribution of the semion trapped in the vison loop as well. Hence we conclude that stacking along three directions gives another $2^3=8$ possibilities (independently) on top of placing background fractional excitations. The stacking of e-SPT can be probed by linking a vortex loop with a dislocation half-plane and consider its projective translation or braiding with another loop as described above.

Another potential bosonic SPT state is the $E_8$ state\cite{kitaev_2006} with a thermal hall conductivity $\kappa_{xy}/T=8\frac{e^2}{h}$.If one imposes a $Z_2$ symmetry of boson number conservation modulo $2$ and gauge it, the $\pi$ flux of the $Z_2$ symmetry is a boson. Furthermore we can regard the $E_8$ state as a combination of the trivial gapped state of the $e$-bosons together with an $E_8$ state formed by $Z_2$ neutral excitations.  Hence stacking gauged $E_8$ states do not really give new SET classes; they merely correspond to the ability to combine the $Z_2$ SET states with invertible states obtained by stacking layers of the $2+1$-d $E_8$ state. We will therefore exclude these from our classification. 

In total we have $16\times 8=128$ $3+1$d SET classes of translation-enriched $Z_2$ topological order.

\section{Spectroscopic signature of translation symmetry fractionalization in the honeycomb vison model}
\label{app:neutron}
\begin{figure}
 \captionsetup{justification=raggedright}
    \centering
        \adjustbox{trim={.24\width} {.4\height} {.14\width} {.08\height},clip}
    {\includegraphics[width=.65\textwidth]{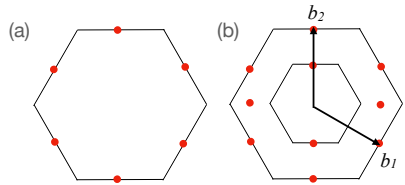}}
    \caption{The momenta at which the enhanced periodicity implies continuum excitations in the spin structure factor (assuming there is such scattering at $\Gamma$)   for pairs of majoranas ($\epsilon$ particles) in (a) and pairs of visons ($e,m$) in (b) respectively. Due to the anti-commuting relation of $T_{1,2}$ for $\epsilon$ and $T_2^2,T_1$ for $e,m$, the neutron spectrum is periodic by a momenta transfer of $(0,\pi),(\pi,0),(\pi,\pi)$ for $\epsilon$ and $(\pi,0),(\pi/2,0)$ (or sum of the two vectors) for $e,m$, respectively\cite{hermele_neutron}. (b) also draws the reduced Brillouin zone and reduced reciprocal vectors $b_i\cdot R_j=2\pi \delta_{ij}(i,j=1,2)$ for the vison model we use to fit $\alpha-$RuCl$_3$. Visons are permuted upon translation $T_2$.}
    \label{fig:neutron}
    \end{figure}
    
As discussed in ref\cite{hermele_neutron}, the projective representation of translations for visons will result in distinct signature as an enhanced periodicity  of the continuum associated with 2-anyon excitations in the spin structure factor. Note that there. are three types of continuum spectra that we can expect, associated with exciting pairs of each of the 3 kinds of anyons. 

Here we generalize the discussion to cases where translation permute $e,m$. For the vison model in the main text fig 2, $e,m$ obeys 
\begin{align}
    \label{eq:psg_em}
    T_1 T_2^2T_1^{-1}T_2^{-2}=-1,
\end{align} and $\epsilon$ obeys 
\begin{align}
\label{eq:psg_ep}
    T_1 T_2T_1^{-1}T_2^{-1}=-1
\end{align} due to background anyons. Consider a state with two visons $e_{1,2}$  far away from each other denoted as $|v\rangle$, and apply $T_1$ to one of the visons. From the PSG of eq \eqref{eq:psg_em}, we have for the momentum along $T_2$
\begin{align}
   2 k_2(|v\rangle)=2k_2(T_1(e_1)|v\rangle)+\pi,
\end{align}
where $T_1(e_1)$ denotes applying $T_1$ locally to $e_1$ only and the factor of 2 appears since $T_2^2$ is used for the PSG of $e$. Similarly applying $T_2^2$ to $e_1$, we have
\begin{align}
   k_1(|v\rangle)=k_1(T_2^2(e_1)|v\rangle)+\pi.
\end{align}
Hence we have for the density of states of the two-particle states of $e$ (at a fixed frequency)
\begin{align}
    N_e(\vec k)=N_e(\vec k+(0,\pi/2))=N_e(\vec k+(\pi,0)),
\end{align}
in the reciprocal vector basis $b_{1,2}$ marked in fig \ref{fig:neutron}(b) and similarly for $m$.

For $\epsilon$ the case is simpler and follow the analysis in ref \cite{hermele_neutron}, we have
\begin{align}
    N_\epsilon(\vec k)=N_\epsilon(\vec k+(0,\pi))=N_\epsilon(\vec k+(\pi,0)).
\end{align}

The periodicity of anyon density of states in momentum space will result in the same period in spin structure factor. The spin operator is generally decomposed to contain vison or $\epsilon$ bilinears at leading order in low energy, e.g. for visons one has 
\begin{align}
\vec S(\vec q)\sim \vec f(\vec q)\sum_k v(k)v(-k+\vec q)+ \vec g(\vec q)\sum_k v^\dagger(-k)v(-k+\vec q)+h.c.,
\end{align}
where $\vec f,\vec g(\vec q)$ are form factors which generically depend on $\vec q$. Hence the spin structure factor probes the two-vison density of states $N_e$ modulated by the form factors. So a period in $N_e$ in momentum space does not lead directly to an exact period in spin structure factor $\langle S(q,\omega) S(-q,-\omega)\rangle$, though one expects the general features in neutron spectroscopy, i.e. peak or continuum, should be repeated qualitatively by a period in vison/$\epsilon$ density of states.

If we assume continuum excitations at the $\Gamma$ point, as observed experimentally, then the other momenta implied by the enhanced periodicity are marked in fig \ref{fig:neutron}. Note any combination of the periodicity momenta marks also a period.

\end{document}